\documentclass[onecolumn,showpacs,showkeys,amsmath,amssymb,superscriptaddress,floatfix,nofootinbib]{revtex4}
\usepackage[utf8]{inputenc}
\usepackage{epsfig,amsmath,amssymb}
\usepackage{bm}
\usepackage{graphicx}
\usepackage{graphicx}
\usepackage{dcolumn,color}
\newcommand{\ave}[1]{\left\langle #1 \right\rangle}
\newcommand{\sfrac}[2]{{\textstyle\frac{#1}{#2}}}
\newcommand\di{\partial}

\newcommand{\order}[1]{ \mathcal{O} \left( #1 \right) }
\newcommand{\eqcomma}{\phantom{AA},\phantom{AA}}
\newcommand{\lnz}{\ln \mathcal{Z}}
\newcommand{\lag}{ \mathcal{L}}

\def\k{{\bf k}}

\begin{document}

\title{Linear response theory and effective action of relativistic hydrodynamics with spin}
\author{David Montenegro$^{1,2}$,Giorgio Torrieri}
\affiliation{IFGW, Unicamp, Campinas,SP, Brazil $\phantom{A}^2$ Instituto de Fisica Teorica, Universidade Estadual Paulista, Sao Paulo, SP, Brazi}
\begin{abstract}
We use linear response techniques to develop the previously proposed
relativistic ideal fluid limit with a non-negligible spin density.
We confirm previous results \cite{gt1,gt2,gt3}, obtain expressions for the microscopic transport coefficients
using Kubo-like formulae and build up the effective field theory from the computed correlation functions.   We verify that for a causal theory with spin the spin-polarization correlator's asymptotic time dependence is the same as for fluctuating hydrodynamics, and investigate backreaction corrections to hydrodynamic variables using a one-loop effective action.
We also confirm that polarization makes
vortices acquire an effective mass via a mechanism similar to the Anderson-Higgs
mechanism in superconductors.
As speculated earlier, this could stabilize the ideal hydrodynamic
limit against fluctuation-driven vortices
  \end{abstract}

\maketitle
\section{Introduction}
An interesting problem in  relativistic fluid dynamics is the inclusion of a non-zero polarization density within the fluid.    This is a challenging problem even at an intuitive level because several characteristics associated with ideal fluids such as isotropy and conservation of circulation, will not apply when spin density is non-zero.
Indeed, several approaches have been tried \cite{gt1,gt2,gt3,flork1,flork2,flork3,gale,hongo,bec,nair,dirk} , with a consensus on even the fundamental dynamics still lacking.

One virtue of the Lagrangian approach \cite{gt1,gt2,gt3} is that it allows us to start from local equilibrium {\em as an assumption} and build up the Lagrangian from the free energy, independently of the underlying microscopic theory.
Essentially, we do not know what the system looks like microscopically but we know that its dynamics is ``strongly coupled and high temperature enough'' that the system quickly adjusts itself to local equilibrium after perturbed.  That ``quickly'' leads to a separation of scales w.r.t. the gradient and timescale of the perturbation.

Of course this ``bottom up'' approach has quite a few limitations.
For example, transport coefficient's dependence on temperature, angular momentum and chemical potential necessitate knowledge of the underlying microscopic theory. 
Nevertheless, bottom-up reasoning has allowed us to obtain several results in an intuitive way, such as the necessity of parallelism between spin and angular momentum \cite{gt1} and the necessity of dissipation for a causal theory \cite{gt2,gt3}. 

In this work, we cement these previous results reformulating hydrodynamics with spin in terms of linear response and correlation functions.  
This develops the Lagrangians of \cite{gt1,gt2,gt3} into a real-bottom up effective theory, explicitly  including the response of the bulk hydrodynamic evolution to microscopic fluctuations and correlations.

Our Lagrangian, following \cite{nicolis,nicolis1,gt0} (which mixes the Keldysh-Schwinger like prescription of \cite{groz,floer} with the ``many particles to continuum'' approach used in \cite{nair,jackiw,comer} )
contains the information of the equation of state and entropy current in terms of the field $\phi_I$ of the Lagrangian coordinates of the fluid element\footnote{Hereafter Greek letters refer to Lorentzian 4D lab coordinates and latin ones to co-moving 3D Euclidean ones}.  The entropy of the volume element is then proportional to the volume of the element 
\begin{equation}
  \label{bdef}
b = \left( \det_{IJ} \left[ \partial_\mu \phi_I \partial^\mu \phi_J \right] \right)^{1/2}
  \end{equation}
in the absence of chemical potentials this is the only propagating degree of freedom possible.   Including  the polarization tensor $y_{\mu \nu}$ is similar to including  a chemical potential which however transforms as a vector in the co-moving frame \cite{gt1,gt2}.   Since spin density is not conserved, $y_{\mu \nu}$ is an auxiliary field interacting with $b$ via the equation of state rather than an extension of $b$ (as the microscopic phase generating the chemical potential is in \cite{nicolis}).
For a well-defined local equilibrium vorticity and polarization need to be parallel  \cite{gt1}, 
\begin{equation}
  \label{parall}
  y_{\mu\nu} = \chi(b,\omega_{\mu\nu}\omega^{\mu\nu})\omega_{\mu\nu}, 
\end{equation}
where $\chi(\omega,b)$ is the vortical susceptibility and the relativistic vorticity \cite{rezzolla} includes the enthalpy $w$ 
\begin{equation}
\label{vortdef}
\omega_{\mu \nu} = 2 \nabla_{[\mu} w u_{\nu]} 
 = 2 w \left( \nabla_{[\mu}u_{\nu]} - \dot{u}_{[\mu} u_{\nu]} +  u_{[\mu} \nabla_{\nu]} \ln w \right) 
\end{equation}
Note that while
$y_{\mu \nu}$ breaks local isotropy explicitly, it not vanish at thermodynamic equilibrium when angular momentum is present \cite{beclisa,becwigner}.

Indeed, in the case where vortical susceptibility $\chi$ is calculated explicitly \cite{gt1}, the expression for magnetic and vortaic susceptibility parallel each other, suggesting the dynamics is the same up to charge conjugation symmetry.
In a fluid with no chemical potential one expects the spin alignment will not
produce a magnetic field (since the magnetic moment of particles and antiparticles is opposite), but it will break isotropy and take angular momentum out of vorticity and vice-versa.

In \cite{gt1} we have shown that, for a ``paravortaic'' equation of state
\begin{equation}
  \label{eos}
F(b,y) = F\left(b(1-c y_{\mu \nu} y^{\mu \nu}) \right)
  \end{equation}
this Lagrangian leads to three conservation law type equations $\partial_\mu J^\mu_I=0$
\[\
 J^\mu_I = 4\, c \, \partial_\nu \left\{ F^\prime \left[ \chi  \left(   \chi + 2 \, \frac{d\chi}{d\omega^2} \right) \omega_{\alpha\beta} \, g^{\alpha\{\mu} P_I^{\nu\}\beta} \right]\right\}  - F^{\prime} \left[ u_\rho P^{\rho\mu}_I \left(   1-c y^2 - 2  c  b  \chi  \omega^2 \, \frac{d \chi}{db} \right)\right] - 2 c \left(   \chi + 2 \, \omega^2 \, \frac{d \chi}{d\omega^2} \right)F^{\prime} \times \] 
\begin{equation}
\label{jdef}
  \times \left\{ \left[ \chi \, \omega^2 -\frac{1}{b}y_{\rho\sigma} \left(   u_\alpha \partial^\alpha (bu)^\rho - u_\alpha \nabla^\rho (bu)^\alpha \right) \right] P^{\sigma\mu}_I   - \frac{1}{6 b} y_{\rho\sigma}\varepsilon^{\mu\rho\alpha\beta}\epsilon_{IJK}\nabla^\sigma\partial_\alpha \phi^J \partial_\beta \phi^K \right\}.
\end{equation}
with the projector $P^{\mu \nu}_K = \partial (bu)^\mu /\partial(\partial^\nu \phi^K)$, $\nabla^\alpha = \Delta^{\alpha \beta} \partial_\beta$ and 
$[...],\{...\}$ corresponding to, respectively, antisymmetrization and symmetrization of the indices, as done in \cite{rombook,kodama}.

This is the ``ideal hydrodynamic limit with polarization'', the equation of motion of a fluid with spin density where local equilibrium is reached instantaneously.

However, as shown in \cite{gt2,gt3} this equation produces non-causal perturbations. causality means equation \ref{vortdef} can only be achieved as a relaxation asymptotic limit,
\begin{equation}
  \label{IS}
\tau_Y \partial_\tau \delta Y_{\mu \nu} + \delta Y_{\mu \nu} = y_{\mu \nu} =  \chi(b,w^2) \omega_{\mu \nu}   
\end{equation}
Equation \ref{jdef}, analogously to other Maxwell-Cattaneo cases, needs to be updated with $Y_{\mu \nu}$ as an additional degree of freedom, the ``magnon''.
However, this theory is still as close to local equilibrium as possible, since the non-equilibrium ``Magnon'' tensor $Y_{\mu \nu}$ just relaxes to the equilibrium value.
Since the magnon as an independent degree of freedom propagates and interacts, any further non-dissipative dynamics for the magnon \cite{flork1,flork2} have the potential to bring the system arbitrarily far away from local equilibrium, and hence can not lead to well-defined effective field theories around the local equilibrium state.

The next three sections will link these results to the more traditional linear response theory.   A correlation function for $J_I^\mu,T_{\mu \nu}$ and $Y_{\mu \nu}$ will be derived.   A fluctuation dissipation relation linking $\chi$ and $\tau_Y$ will also be derived.

Using these techniques, we are able to confirm and develop several novel results.    We show that polarization's breaking of local isotropy can be characterized by a dynamics in some ways analogous (although dissipative) to the Anderson-Higgs mechanism in Gauge theory, with the polarization condensate giving an effective ``mass'' (through a dissipative imaginary one \cite{massdiss}) to the vortex.
Thus, polarization realizes Landau's original observation \cite{landau} that to stabilize hydrodynamics vortices must have a mass gap (this turned out to describe superfluidity, but not ordinary fluids. In contrast, \cite{nicolis1} conjectured that there is no stable ''quantum'' theory of fluids and \cite{gripaios} argued that for such a theory to exist only conserved observables are allowed).

We also verify that  vorticity-polarization correlation exhibits the same long time-tail behaviour that characterizes the Maxwell-Cattaneo equation fluctuations examined in \cite{kovtun,kovtunlec}.   And we use effective action techniques to calculate the backreaction of sound waves and vortices on hydrodynamic variables.

We should reiterate that the results here come exclusively from ``bottom-up'' reasoning, independently from the microscopic theory:  We assume we are close to local equilibrium, and use gradient expansions, causality, and unitarity analysis for derivation, together with the results of \cite{gt1,gt2,gt3}.    This is in contrast with most approaches \cite{flork1,flork2,flork3,gale,dirk} which rely on a ``top-down'' microscopic description, usually using extensions of the Boltzmann equation.    Eventually, the two approaches should of course be verified to coincide, via the matching of the free energies and transport coefficients calculated here with those calculated in the microscopic models, but at the moment we are far away from this.
One reason is that the different approaches are not yet consistent, with some admitting an RTA expansion \cite{flork3} and others casting doubt this is possible \cite{dirk,flork1}, some admitting a gradient expansion and some where some gradients explicitly diverge \cite{flork1}.   Furthermore, expansions assuming a spinorial Wigner function necessarily miss the admixture of particles with different spin.

Nevertheless, there are tantalizing hints the tow approaches can indeed be reconciled. For instance, a very similar conclusion to that discussed in \cite{gt3} and section \ref{reltimeg} can be reached from a purely microscopic calculation \cite{dirk}, where it is shown that the collision term when spin is included necessarily becomes non-local, precluding a causal instantaneous local vanishing of the collision term.

We shall therefore proceed with a bottom-up linear response analysis, but keep in mind the microscopic results for further work.
\section{Linear response analysis \label{pqpqp}}
We  start with the assumption of local equilibrium and linear response theory.   While hydrodynamics is highly non-linear, one assumes that any microscopically-driven perturbation starts off in the linearized stage from the equilibrium state, hence its growth rate can be approximated by a linear response function.    This is also equivalent to assuming these changes develop slowly enough to be considered ``adiabatic'', so local thermalization can be assumed at any moment in its evolution.
Then, the coefficient of the constitutive relation are given by taking the low energy and long wavelength limit for the correlation function of operators in different points in the space and time. Since the system should not, in this regime, distinguish whether the deviation came from either an external disturbance or natural fluctuation, the transport coefficient for linear response and auto-correlation functions must be related.  This is the basis for the relations generally known as ``fluctuation-dissipation theorem'', which here we apply to spin-vorticity dynamics. 
\subsection{The magnon field $Y_{\mu \nu}$ and it's vortex source}
In this spirit, building on \cite{gt1,gt3} let the vorticity-fluctuation coupling be given by the interacting picture Hamiltonian $H_I$ \cite{tong}

\begin{equation}
H_I(t) = \int d^3x Y^{\mu\nu}(t, \vec{x}) \boldsymbol \omega_{\mu\nu} (t, \vec{x})
\end{equation}

where the vorticity $\omega_{\mu\nu}$ is treated as a classical source to the
hermitian polarization operator $Y^{\mu\nu}(t, \vec{x})$. The unitary operator generating the temporal evolution reads

\begin{equation}
    U(t,t_0) = \mathcal{T}(e^{-i \int^t_{t_0} H_I(t') dt' })
\end{equation}

Where $\mathcal{T}$ is a time-ordering product. To understand how $\boldsymbol \omega_{\mu\nu}$  affects the field operator $Y^{\mu\nu}$ as a deviation from local equilibrium, we look at the density matrix in Heisenberg picture

\begin{equation}
\rho = U(t,t_0) \rho_0 U^\dagger(t,t_0), \qquad \rho = \frac{1}{Z} \sum_\alpha e^{-\beta H_\alpha - \vec{\omega} \cdot \vec{J} / T }
\end{equation}
 
Linearizing from the equilibrium expectation value we get

\begin{equation}\begin{aligned}
\label{ymunueq}
\langle Y^{\mu\nu} (t, \vec{x}) \rangle_{\omega} 
= \langle \rho_0 (t) U^\dagger(t,t_0) Y^{\mu\nu} (t, \vec{x}) U(t,t_0) \rangle \approx
\end{aligned}\end{equation}
\[\  \approx \langle Y^{\mu\nu}(t, \vec{x}) \rangle_{\omega_{\alpha \beta}=0} +  i \int dt^\prime  \int d^3 x^\prime \langle [Y^{\mu\nu}(t,\vec{x}),Y^{\alpha\beta}(t^\prime,\vec{x}^\prime)] \rangle_{eq}  \boldsymbol\omega_{\alpha\beta}(t^\prime, \vec{x}^\prime) \]

with $\langle  \delta Y^{\mu\nu}  \rangle = \langle Y^{\mu\nu} (t, \vec{x}) \rangle_{\omega} - \langle Y^{\mu\nu}(t, \vec{x}) \rangle_{\omega_{\alpha \beta}=0}$. where $\langle Y^{\mu\nu}(t, \vec{x}) \rangle_{\omega_{\alpha \beta}=0}$ is classical average at local equilibrium, before the vortex (source) switched on. The induced polarization reads

\begin{equation}
\langle  \delta Y^{\mu\nu}  \rangle  = i \int^{+ \infty}_{- \infty} dt^\prime  \int d^3 x^\prime e^{\epsilon t^{\prime}} \Theta(t -t^{\prime}) \langle [Y^{\mu\nu}(t,\vec{x}),Y^{\alpha\beta}(t^\prime,\vec{x}^\prime)] \rangle_{eq}  \boldsymbol\omega_{\alpha\beta}(t^\prime, \vec{x}^\prime)
\end{equation}

In order to arrive at Kubo's formula we assume an adiabatic change in local vorticity

\begin{equation}\begin{aligned}\label{xxxx}
\boldsymbol\omega^{\mu\nu}(\vec{x}^{\prime}, t^{\prime}) &= e^{\epsilon t^{\prime}} \Theta(-t^{\prime}) \boldsymbol\omega^{\mu\nu}(\vec{x}^{\prime}), \qquad t<0 \\
\boldsymbol\omega^{\mu\nu}(\vec{x}^{\prime}, t^{\prime}) &= 0, \qquad \ \ \ \ \ \ \ \ \ \ \ \ \ \ \ \ \ \ \ \ \ \ t>0
\end{aligned}\end{equation}

The adiabatic assumption allows us to add a factor $e^{\epsilon t}$, which smears out any short-range fluctuations. In the physical sense, it ensures a smooth evolution of spin-orbit balance until the transport process archives thermalization. In addition, in the absence of spontaneous symmetry breaking (see conclusion of \cite{gt3}) the $\frac{1}{\epsilon}$ conducts the transition rate in which the spins interacting with each other (later in section \ref{reltimeg} we will see how to correct this prescription to ensure causality using the relaxation time).   In this case, the linear approximation  remains valid and the physical meaning of $\epsilon$ in the source in Eq. \ref{xxxx}  is a rate of increasing $\boldsymbol \omega^{\mu\nu}$, or density of spin aligned with vortex. Applying the equation \ref{xxxx}, we get the expectation value of correlation function between two operator in different point of space-time within a system in equilibrium

\begin{equation}\begin{aligned}\label{fd}
\langle  \delta Y^{\mu\nu}  \rangle  &= i \int^t_{- \infty} dt^\prime  \int d^3 x^\prime e^{\epsilon t^{\prime}} \Theta(t -t^{\prime}) \langle [Y^{\mu\nu}(t,\vec{x}),Y^{\alpha\beta}(t^\prime,\vec{x}^\prime)] \rangle_{eq}  \boldsymbol\omega_{\alpha\beta}(t^\prime, \vec{x}^\prime), \ \ \ t<0 \\
&= i \int^0_{- \infty} dt^\prime  \int d^3 x^\prime e^{\epsilon t^{\prime}} \Theta(t -t^{\prime}) \langle [Y^{\mu\nu}(t,\vec{x}),Y^{\alpha\beta}(t^\prime,\vec{x}^\prime)] \rangle_{eq}  \boldsymbol\omega_{\alpha\beta}(t^\prime, \vec{x}^\prime), \ \ \ t>0
\end{aligned}\end{equation}

The commutator above expresses the retarded green function defined as

\begin{equation}\begin{aligned}
G^R_{Y^{\mu\nu}Y^{\alpha\beta}} (\vec{x}^{\prime},  \vec{x} , t^{\prime}, t )  \equiv -i \Theta( t - t^{\prime}) \langle 0 [Y^{\mu\nu} (\vec{x},t), Y^{\alpha\beta} (\vec{x}^{\prime},t^{\prime})]  0 \rangle_{eq}
\end{aligned}\end{equation}                      

The polarization-polarization correlation function gives the amount of fluctuation correlated in space and time and the Heaviside step function $\Theta(t-t^{\prime})$ assures causality in all reference frames provided the dispersion relation is subluminal.
We shall rewrite the eq. \eqref{fd} as that of a single ``Retarded'' Greens function $G^R$

\begin{equation}
\langle  \delta Y^{\mu\nu}  \rangle = i \int^0_{- \infty} dt^\prime  \int d^3 x^\prime e^{\epsilon t^{\prime}} G^R_{Y^{\mu\nu}Y^{\alpha\beta}} (\vec{x}^{\prime},  \vec{x} , t^{\prime} - t )  \boldsymbol\omega_{\alpha\beta}(t^\prime, \vec{x}^\prime), \ \ \ t>0
\end{equation}
This derivation makes explicit the connection between the  transport coefficient and the  ``hydrodynamic pole'' which in $\chi=0$ hydrodynamics has a dispersion relation and determines the sound modes \cite{kovtunlec}.  The next subsection will use this to calculate susceptibility.

\subsection{Susceptibility \label{chisec}}
Susceptibility, $\chi(b,\omega^2)$ in Eq. \ref{parall}, tell us how polarization behaviour is induced from small perturbation around equilibrium and after turning off $\omega^{\alpha \beta}$. We obtain the susceptibility by taking the hydrodynamic regime, where the comoving coordinates are perturbed against the hydrostatic coordinates $X_I$.   
\begin{equation}
  \label{staticdef}
  \phi_I = X_I + \pi_I 
\end{equation}
given a wavenumber $k_I$ $\pi_I$ can be separated into a sound-wave and a vortex part
\begin{equation}
  \label{soundvort}
\pi_I= \pi_I^T(k_I)+\pi_I^L(k_I) \eqcomma  k_I \pi_I^T=0 \eqcomma k_I \pi_I^L = |k||\pi|
  \end{equation}
then the susceptibility becomes
\begin{equation}
 \chi = \lim_{k\rightarrow 0} \bigg( [ \langle \pi^T_a \pi^T_b \rangle - \langle \pi^T_a \rangle \langle \pi^T_b \rangle ] + [ \langle \pi^L_a \pi^L_b \rangle - \langle \pi^L_a \rangle \langle \pi^L_b \rangle ] \bigg) / \mathcal{V}
\end{equation}
where $\mathcal{V}$ is the volume of phase space in the thermodynamic limit and $\chi_{ab}$ is symmetric matrix ($ det (\chi) \geq 0$) whose diagonal form is realized when aligned with the rotation axes. 
The limit of small frequency

\begin{equation}\begin{aligned}
\chi(b,\boldsymbol\omega^2) &= \frac{\delta Y^{\mu\nu} }{\delta \omega^{\alpha\beta}}\bigg|_{ \boldsymbol\omega=0} \delta_\alpha^\mu \delta_\beta^\nu
  \end{aligned}
\label{chideriv}
\end{equation}  

where $\chi(b,\boldsymbol\omega^2)$ (thermodynamic derivative) is a statistic thermodynamic quantity.  The susceptibility is an analytical function, which the poles lies below the real axis. Thus we may split it

\begin{equation}
\chi (\omega + i \epsilon , \vec{k} ) = \chi^{\prime} ( \omega , \vec{k}) + i \chi^{\prime \prime} (\omega , \vec{k})    
\end{equation}
in an imaginary $\chi^{\prime \prime}$ (absorptive or dissipative) and real $\chi^{\prime}$ (symmetric in time) parts.  

The relation of imaginary part of susceptibility  with retarded green function

\begin{equation} 
\langle 0 [Y^{\mu\nu} (\vec{x},t), Y^{\alpha\beta} (\vec{x}^{\prime},t^{\prime})]  0 \rangle_{eq} = \int d\omega d \vec{k}  \frac{\chi^{''}(\omega,\vec{k})}{\omega} e^{i( \omega (t-t') - \vec{k}(\vec{x}-\vec{x'}))} 
\end{equation}

provides a link between a linear response described by a linearized hydrodynamic and correlation function. The $\chi^{\prime \prime}$ is responsible for all information on commutator of polarization. From the definition of Green function, we have the Kramers-Konig relation \cite{tong} linking the retarded Greens function $G^R$ to the advanced one $G^A$

\begin{equation}
Re[G^R] = Re[G^A] \qquad, \ \ \ Im[G^R] = - Im[G^A]
\end{equation}

To proceed further we look at
the spectral representation

\begin{equation}G^{R/A} = \int \frac{d \omega^\prime}{2 \pi} \frac{ \rho^{Y Y'} (\omega, \vec{k} ) }{\omega^\prime - \omega \mp i \epsilon}\end{equation}              

The spectral density $\rho^{Y Y'}$ is non negative and a real function containing the density of state at frequency $\omega$ and $\vec{k}$.

Let us discuss the properties of correlation function $\langle \phi_I (0) \phi_K (t) \rangle = \langle \phi_K (0) \phi_I (-t) \rangle$.  These functions are independent of detailed configuration of the system unless an external field is applied.  The symmetry under time reversal manifests in the form of $\langle \phi_I (t) \phi_K (t^\prime) \rangle = \langle \phi_I (t^\prime) \phi_K (t) \rangle$. In our picture of fluctuation, we have to know how the fields changed under reverse of velocity due to switch direction of external rotating frame from $\Omega$ to $-\Omega$. From this perspective, any even combination of longitudinal ($\pi_L$) or transverse ($\pi_T$) are unaffected by reversal $\Omega$ and thus the retarded green function exhibits the following property 

\begin{equation}
G^R_{ab} (t, \Omega) = G^R_{ba} (t, - \Omega) = G^R_{ab} (- t, - \Omega) = G^R_{ab} ( - t, \Omega), 
\end{equation}

The general formulation is $G^R_{ab}(\omega, \vec{k}, \Omega) = \eta_a \eta_b G^R_{ab}(\omega, \vec{k}, \Omega) $, where $S = diag (\eta_1,\eta_2, \dots) =  diag(1,1,\dots)$
whereas the kinetic coefficients takes the form

\begin{equation}
\gamma_{ik}(\Omega) = \gamma_{ki}(-\Omega)
\end{equation}

The Hamiltonian under time reversal operator reads

\begin{equation}
\Theta H(\omega)\Theta^{-1} = H(-\omega)    
\end{equation}

If we simultaneously change the signal of all charge, the current and $\Omega$ remain in the same direction. There is no charge conjugation so in Eq. \ref{jdef} $J_I^\mu \rightarrow J_I^{\mu *}$ and up to first order it is easy to see that $J_I^\mu \sim k^\alpha \omega_{\alpha I} \delta^\mu_0$ is consistent with this symmetry. To develop the necessary tool for the evaluation of average at disturbed system, we rewrite the locally conserved Noether current at a variational principle ( as is done in the next section \ref{corrfunc}), but for now we know that the
`` conserved current'' of Eq. \ref{jdef} are generated by $\hat{w}^\alpha = \epsilon^{\alpha \mu\nu}\omega_{\mu\nu}$ and $\omega^{\mu\nu}$ the vorticity field.


We want to investigate how the gradient of hydrodynamical variable disappear by an external disturbance. This allows us to introduce the systematic of Kubo formulae for current. The general linear transformation reads

\begin{equation}\begin{aligned}
& \langle J_I^\mu (t,\vec{x})  \rangle_{| \omega} - \langle J_I^\mu (t,\vec{x}) \rangle_{0} = \int d^3 x^{\prime} d t^{\prime}  \langle [J_I (t,\vec{x}),J_I (t^{\prime},\vec{x}^{\prime})] \rangle \hat{\omega}^{\mu} \\ \\
& \langle J_I^\mu (t,\vec{x}) \rangle_{0} =   2 \, c \, \vphantom{\frac{}{}} \chi^2  F^{\prime} (b_0) b_0 \, \boldsymbol\omega^2 \delta^\mu_I  \\ \\ 
& \langle J_I^\mu (t,\vec{x})  \rangle_{| \omega} = 2 \, c \, \vphantom{\frac{}{}} \chi^2  F^{\prime} \, \boldsymbol\omega^2 \delta^\mu_I  +  \frac{\chi_{ij}^{\prime \prime}(b, \boldsymbol\omega^2  ) }   {i \omega}
\end{aligned}\end{equation}

The main consequence of Eq. \ref{chideriv} is that $\langle Y^{\mu\nu}  \rangle_{\omega_{\alpha \beta}=0}$ refers to the microscopic quantum operator average before the background field is slowly switched on.

Looking at Eqs \ref{staticdef} and \ref{soundvort}, 
in the limit when $\chi=0$ \cite{nicolis1}, an applied vortex has a similar form as the magnetic field potential in electromagnetic theory.  Although the vortex has no propagation, we can treat the flow as a gauge covariant derivative which performs an infinitesimal rotation in each fluid cell. It is easy to see how local translational symmetry connects with Euler and Lagrangian picture, thus following this line of thought, the lagrangian density appears invariant under local $SO(3)$ symmetry.  So long as there is no dissipation, the vortex behaves exactly as a gauge field, with the ocovariant derivative ``propagating'' the gauge element along the fluid.  This symmetry also keeps the vortex massless \cite{nicolis1}.

In this respect, as shown in more detail in section \ref{seclint}, polarization acts as a Higgs mechanism \cite{higgs} giving the vortex a ''mass'' related to $\chi$.
The non-vanishing of the vacuum expectation value of classical field is similar to the Higgs mechanism, in that we can roughly compare the ``mass'' particles get by interacting with the Higgs expectation value with the ``dissipative mass''  $  M^2 = ( \chi^2\sqrt{F^{\prime}(b_0) b_0})^{-1}$ where $w_0$ is the background enthalpy\footnote{Unfortunately it is standard to refer to $w$ as enthalpy as well as frequency, and $\omega_{\mu \nu}$ to vorticity.   In most of our paper we only use $w_0$ for background enthalpy and $w$ within Fourier integrals for time frequencies.  The reader should nevertheless be careful with the context of each expression } . One can use the background value of Lagrangian $ F_{b}(b_0 = 1) = w_0$ and consider that $w_0$ multiply every term in the current. The effective mass from dissipative interaction between the spin and fluid volume element under the presence of rotation is $M^2 = ( \chi^2)^{-1}$. The residual part of $\hat{\omega}^\mu$ is not an elementary field but an excitation, and plays a role of source (this will be examined in detail in section \ref{interhyd} ).

Unlike the usual Higgs mechanism, however, Gauge symmetry does not allow us to remove Ostrogradski's instabilities even in the linearized limit \cite{gt2}.  Dissipation therefore becomes necessary \cite{gt3}.   In the section \ref{reltimeg}  we shall see the effect it will have in correlation functions, but first we need to comute the complete correlation structure of the non-dissipative theory. 
For this, functional methods will become important.
\section{Correlation functions from functional methods \label{corrfunc}}
In the previous section, we saw that the Green-Kubo relation can be defined by hydrodynamic variable which attained the transport equation from conservation law.   However, hydrodynamics with polarization cannot \cite{gt1} entirely be written in terms of such laws.      To get the more general correlation functions allowed within the theory it is easier to use functional methods.

In this approach, one can define a partition function $\lnz$ so that average quantities are derivatives w.r.t. generators.  We know that the energy momentum tensor is generated by gravitational tensor perturbations $h^{\mu \nu}$ and the conservation currents in Eq. \ref{jdef} are generated by $\hat{w}^\alpha = \epsilon^{\alpha \mu\nu}\omega_{\mu\nu}$
\begin{equation}
  \label{lnzt}
  \mathcal{T}^{\mu\nu} (x) \equiv \sqrt{-g} \langle T^{\mu\nu} (x) \rangle_{\omega,g} = \left. \frac{\delta}{\delta h^{\mu \nu}} \lnz \right|_{h^{\alpha \beta}=\hat{\omega}^\alpha=0}
\end{equation}
\begin{equation}
\label{lnzj}
  \left. \mathcal{J}^\mu (x) \equiv \sqrt{-g} \langle J^\mu (x) \rangle_{\omega,g}= \frac{\delta}{\delta \hat{w}_{\mu }} \lnz \right|_{h^{\alpha \beta}=\hat{\omega}^\alpha=0}
\end{equation}
We shall characterize long-distance dynamics properties by a small fluctuation from off-diagonal metric around a static background.  We are essentially updating, in the polarization context discussed in \cite{gt3}, the analysis made in \cite{kadanoff} and applied to relativistic hydro in \cite{kovtun,kovtunlec}.  The green`s functions of the conserved currents, in our case the energy momentum tensor and the $J^\mu_I$ will be, by the fluctuation dissipation theorem, the {\em second} derivatives of $\lnz$.  Hence
\begin{eqnarray}\label{eq:GR-sources}\begin{aligned}
G^{\, R}_{J^\mu \! J^\nu}(x) & = -
\left.\frac{\delta {\cal J}^{\mu}(x) }{\delta \hat{\omega}_{\nu}(0)}\right|_{h^{\alpha \beta}=\hat{\omega}^\alpha=0}\,, &
G^{\,R}_{T^{\mu\nu} \! J^\sigma}(x) & = -
	\left.\frac{\delta {\cal T}^{\mu\nu}(x)}{\delta \hat{\omega}_{ \sigma}(0)}\right|_{h^{\alpha \beta}=\hat{\omega}^\alpha=0}\,, \\[7pt]
	G^{\,R}_{J^\sigma T^{\mu\nu}}(x) & = - 2
	\left.\frac{\delta {\cal J}^{\sigma}(x) }{\delta h_{\mu\nu}(0)}\right|_{h^{\alpha \beta}=\hat{\omega}^\alpha=0}\,, &
	G^{\,R}_{T^{\sigma\tau} T^{\mu\nu}}(x) & = - 2
	\left.\frac{\delta {\cal T}^{\sigma\tau}(x) }{\delta h_{\mu\nu}(0)}\right|_{h^{\alpha \beta}=\hat{\omega}^\alpha=0}\,.
\end{aligned}
\end{eqnarray}

The covariant derivative provide the interplay between out-of-equilibrium hydrodynamic variable and the metric and vorticity sources, with a vector and a tensor term (the latter can be rewritten as a ``graviton'' $h_{\mu \nu}$)

\begin{equation}
    \nabla_\mu u^\nu = \partial_\mu u^\nu + \frac{1}{2} \eta^{\nu \beta} (\partial_\mu h_{\beta \rho} + \partial_\rho h_{\beta \mu} - \partial_\beta h_{\mu \rho})u^\rho
\end{equation}
Beyond leading order, rather than the metric form of Eq. \ref{lnzt}, we shall use the canonical non-symmetrized tensor, as one can explicity calculate it in terms of $\phi_I$ (summation in indices $I$ of $\phi_I$ and $J^\mu_I$ omitted for brevity) 
\begin{eqnarray}\label{st}
T^{\mu\nu} &=& \bigg\{ \frac{ \partial \lnz }{\partial (\partial_\mu \phi )  } - \partial_\beta \frac{ \partial \lnz }{  \partial (\partial_\mu \partial_\beta  \phi )  } +  \partial_\beta  \partial_\gamma \frac{ \partial \lnz }{\partial (\partial_\mu \partial_\beta  \partial_\gamma \phi )  } - \dots  \bigg\}   \partial^\nu \phi   + \nonumber \\
&& \bigg\{  \frac{ \partial \lnz }{  \partial (\partial_\mu \partial_\beta  \phi )  } -  \partial_\gamma \frac{ \partial \lnz }{\partial (\partial_\mu \partial_\beta  \partial_\gamma \phi )  } + \dots  \bigg\} \partial_\beta \partial^\nu \phi  +  \bigg\{   \frac{ \partial \lnz }{\partial (\partial_\mu \partial_\beta  \partial_\gamma \phi )  } - \dots  \bigg\} \partial_\beta  \partial_\gamma \partial^\nu \phi + \dots - \eta^{\mu \nu} \lnz
\end{eqnarray}
we note that we use the Canonical rather than the symmetric (Belinfante-Rosenfeld) form of the tensor to keep track of the diffeomorphism-dependent components that couple to $y_{\mu \nu}$ \cite{brauner}.   The integral $\int d^3 x T^{0 i}$ will of course be  independent of pseudo-gauge transformations as expected at the level of the partition function \cite{becwigner}.

The conserved current from Noether theorem of a space-time or internal symmetries for higher order fields is

\begin{equation}
J^\alpha = i \epsilon \left\{ \frac{ \partial \lnz }{\partial (\partial_\mu \phi )  } - \partial_\beta \frac{ \partial \lnz }{  \partial (\partial_\mu \partial_\beta  \phi } + \dots  \right\} \phi   +   \left\{  \frac{ \partial \lnz } {  \partial (\partial_\mu \partial_\beta \phi )}  + \dots  \right\} \partial^\beta \phi 
\end{equation} 

Given the equations \ref{bdef},\ref{parall},\ref{vortdef} and \ref{eos}
one can derive the canonical tensor as
\begin{equation}\begin{aligned}
T^{\mu\nu} &= F^{\prime} \bigg\{ u_\rho P^{\rho\mu}_I  \left( \vphantom{\frac{}{}} 1-c y^2 - 2 \, c \, b \, y_{\alpha\beta} \omega^{\alpha\beta} \, \partial_b \chi \right) \bigg\}  \textcolor{black}{\partial^\nu \phi^I } + \\
&  2 c F^{\prime} (\chi + 2 \omega^2 \partial_{\omega^2} \chi ) \bigg[  \left[ \chi \, \omega^2 u_\sigma -\frac{1}{b} y_{\rho\sigma} \left( \vphantom{\frac{}{}} \dot K^\rho - u_\alpha \nabla^\rho K^\alpha \right) \right] P^{\sigma\mu}_I - \frac{1}{6 b} y_{\rho\sigma}\varepsilon^{\mu\rho\alpha\beta}\epsilon_{IJK}\nabla^\sigma\partial_\alpha \phi^J \partial_\beta \phi^K \bigg] \textcolor{black}{\partial^\nu \phi^I } \\ 
& - \textcolor{black}{\partial_\beta} \bigg( 4\, c\, F^\prime \,\chi  \left( \vphantom{\frac{}{}} \chi + 2 \, \omega^2 \partial_{\omega^2}\chi \right) \omega_{\rho \gamma} \, g^{\rho \{\mu} P_I^{\beta \} \gamma} \bigg) \textcolor{black}{\partial^\nu \phi^I } + \bigg\{ 4\, c\, F^\prime \,\chi  \left( \vphantom{\frac{}{}} \chi + 2 \, \omega^2 \partial_{\omega^2}\chi \right) \omega_{\alpha\beta} \, g^{\alpha\{\mu} P_I^{\nu\}\beta}  \bigg\}  \textcolor{black}{\partial_\beta \partial^\nu \phi^I } - \eta^{\mu \nu} \lnz \label{tmununotsym}
\end{aligned}\end{equation}

and the current, we reproduce the earlier result, the conserved current of Eq. \ref{jdef}


\subsection{Stress tensor perturbations}

The basic idea is expand the variational of stress and current of the eqs. \ref{eq:GR-sources} in terms of the following projectors
\begin{equation}
g^{\mu\nu} = \eta^{\mu\nu} + h^{\mu\nu}, \quad u_\mu \simeq \delta^\mu_0 \left( 1+ \frac{1}{2}\dot \pi^2 \right) + \delta^\mu_I \left( \vphantom{\frac{}{}} -\dot \pi^I + \dot \pi\cdot\partial\pi^I  \right) , \quad 
\omega^2 \simeq  - (\partial_\mu \dot \pi)\cdot(\partial^\mu \dot \pi) - [\partial\dot \pi \cdot \partial \dot \pi]
\end{equation}
  and use the ward identities \cite{boulware,kaja} from energy-momentum conservation
\begin{equation}
\label{warddef}
  \partial_0 G^{\,R}_{ \, T^{00} \, T^{00} }(x) - \partial_i G^{\,R}_{ \, T^{0i} \, T^{00} }(x)  = - \epsilon \omega, \qquad \partial_0 G^{\,R}_{ \, T^{00} \, T^{0i} }(x) - \partial_j G^{\,R}_{ \, T^{0j} \, T^{0i} }(x)  = k_i p
\end{equation}
 to derive all components from one, which comes from the variational principle for the principal modes.    
\begin{equation}
  \label{gencort}
  G^{\,R}_{ \, T^{ij} \, T^{kl} }(x) = - 2 \left.\frac{\delta {\cal T}^{ij}(x) }{\delta h_{kl}(0)}\right|_{h^{\alpha \beta}=\hat{\omega}^\alpha=0}\eqcomma G^{\,R}_{ \, T^{ij} \, T^{kl} }(w,k) = \int d^4 x e^{iwt-kx} G^{\,R}_{ \, T^{ij} \, T^{kl} }(x)
\end{equation}
(thereafter in this and the next section going from $G(x) \rightarrow G(w,k)$ assumes this Fourier transform).
applying this we find
\begin{equation}\label{example}
G^{\,R}_{ \, T^{00} \, T^{00} }(w,k)  =  w_0 \frac{  i  \omega k^2  - i \chi^2   \omega^3 k^2 + i \chi^4   \omega^5 k^2 }{ \chi^2  (\omega^4 + k^2 \omega^2) -  \omega^2 + c_s^2 k^2 }  + i 2 w_0 \chi^2 \omega k^2
\end{equation}

while the other correlators, as expected, agree with the Ward identity Eq. \ref{warddef}

\begin{equation}
G^{\,R}_{ \, T^{00} \, T^{0z} }(w,k)  = w_0 \frac{  i \omega^2 k - i \chi^2   \omega^4 k + i \chi^4   \omega^6 k }{\chi^2   (\omega^4 + k^2 \omega^2) -  \omega^2 +  c_s^2 k^2 }  + i 2 w_0 \chi^2 \omega^2 k
\end{equation}

\begin{equation}
G^{\,R}_{ \, T^{0x} \, T^{00} }(w,k) = 0
\end{equation}

\begin{equation}\begin{aligned}
G^{\,R}_{ \, T^{0x} \, T^{0x} }(w,k) &  = w_0 \frac{ - 2 i \omega^3 - 2 i \chi^4   (\omega^7 - \omega^5 k^2 + \omega^3 k^4 ) + 2 i \chi^2   (\omega^5 - \omega^3 k^2 ) }{ \chi^2   (\omega^4 - k^2 \omega^2) -  \omega^2} - 2 i w_0 \omega  + 2 i  w_0 \chi^2 ( \omega^3 - \omega k^2 / 2)  
\end{aligned}\end{equation}

If $\chi \rightarrow 0$, we will recover the correlators derived previously for the theory without polarization \cite{kovtunlec}, with the finite $\chi$ limit being

\begin{equation}
G^{\,R}_{ \, T^{xy} \, T^{xy} }(w,k)  =  i w_0 \chi^2   \omega^3     
\end{equation}
which also respects the null energy condition
\begin{equation}
\label{null}
  G^{\,R}_{ \, T^{0x} \, T^{0x} }(x) > G^{\,R}_{ \, T^{yx} \, T^{yx} }(x)
\end{equation}

The contact terms come from the $-\eta^{\mu\nu} \lnz$ of stress tensor. It is easy to see if we recalled that the pressure and energy density expanding up to first order given us  

\begin{equation}
\epsilon = - F(b_0) - b_0 F_b(b_0) [\partial \pi] + \mathcal{O}(\pi^2) , \qquad p = F(b_0) -  b_0 F_b(b_0) + [\partial \pi] F_{bb}(b_0) + \mathcal{O}(\pi^2)
\end{equation}
The contribution of the backreaction terms, $\order{\pi^2}$ will be explored in more detail in section \ref{interhyd}
\subsection{Current and vorticity perturbations}

The green function of an energy momentum current can be obtained by a perturbation of the metric ( fixed by the Ward identity w.r.t other)
\begin{equation}
G^{\,R}_{ \, J^{i} \, T^{jk}}(x) = - 2
\left.\frac{\delta {\cal J}^{i}(x) }{\delta h_{jk}(0)}\right|_{h^{\alpha \beta}=\hat{\omega}^\alpha=0}
\end{equation}
so

\begin{equation}
\label{gengrjt}
  G^{\,R}_{ \, J^{0} \, T^{0z}}(w,k)  =  w_0 \frac{ 2 i \chi^4 ( \omega^6 k + \omega^5 k^2 ) - i \chi^2 ( 6 \omega^5  + 2 \omega^3 k^2  - \omega^4 k - 2 \omega^3 k^2 c_s^2  ) - 2 i c_s^2 \omega k^2 }{ \chi^2 (\omega^4 + k^2 \omega^2) - \omega^2 + c_s^2 k^2}
\end{equation}
 We need to clarify that the '' interaction '' of vorticity field with the current arises by the coupling on the plane of vortex instead of the axial vector $\hat{w}^\mu = \epsilon^{\mu \alpha \beta} \omega_{\alpha \beta}$. 

Equations \ref{gengrjt} can be combined with the Ward-like identiy coming from the conservation of both $J$ and the energy-momentum tensor

\begin{equation}
\partial_0 G^{\,R}_{ \, J^{0} \, T^{00}}(w,k) 
-\partial_i G^{\,R}_{ \, J^{0} \, T^{0i}}(w,k)=0
\end{equation}
and the useful formulae
\begin{equation}
    G^{\,R}_{J^z \! \omega^{xz}}(w,k) = - G^{\,R}_{J^z \! \omega^{zx}}(w,k), \qquad G^{\,R}_{J^i \! \omega^{ij}}(w,k) < G^{\,R}_{J^i \! \omega^{0i}}(w,k)
\end{equation}
to obtain infromation about all the vorticity field correlators.  Given that a generic propagator is
\[\  G^R_{J^j \omega^{jk}}(x)= \left. \frac{\delta {\cal J}^{i}(x) }{\delta \omega^{jk}(0)}\right|_{h^{\alpha \beta}=\hat{\omega}^\alpha=0}  \]
the following table gives the full list of components
\begin{equation}
  \begin{array}{|c|c|c|c|c|c|c|c|c|c|}
    \mathrm{Propagator} &
     J^z \! \omega^{0z} &
     J^0 \! \omega^{0z} &
    J^z \! \omega^{0z} &
    J^x \! \omega^{0x} &
    J^0 \! \omega^{0x} &
    J^x \! \omega^{xz} &
    J^z \! \omega^{xz}  &
    J^\mu \! \omega^{\nu \sigma} \\
    &&&&&&&&\\
     \frac{G^R}{w_0 \chi^2} &
      \frac{2i \omega^3 + 2 i \omega k^2 c_s^2 }{\omega^2 - c_s^2 k^2} &
      \frac{4 i k^2 \omega - 4 i c_s^2 k^2 \omega}{\omega^2 - c_s^2 k^2}  &
      \frac{8 c_s^2 \omega^2 k}{\omega^2 - c_s^2 k^2} &
      -6  i \omega   &
     0 &
   \frac{4 i \omega k^2 + 2 i \omega^3  }{\omega^2}  & 
 2 i  \omega &
 0\\
    &&&&&&&&\\
 J^\mu_I & 0 & z & z & 0 & z & x & x & Any
\end{array}
\end{equation}
  This last column is related to the no-anomaly condition, which might be broken in theories such as \cite{surowka} combining local equilibrium with chiral magnetic and vortaic effects.

Similarly, the Ward identity for vorticity

\begin{equation}
\partial_0 G^{\,R}_{ \, T^{00} \, \omega^{0z} }(x) - \partial_i G^{\,R}_{ \, T^{0i} \, \omega^{0z} }(x) = 0
\end{equation}

yields the stress-vorticity correlators

\begin{equation}
G^{\,R}_{ \, T^{00} \, \omega^{0z} }(w,k) = w_0 \frac{ 2 i \chi^2    \omega k^2  + i \chi^2   k \omega^2 - c_s^2 \chi^2   k^3}{\omega^2 - c_s^2 k^2}  
\end{equation}
Taking the Fourier transform of this quantity yields the characteristic ``$t^{-\alpha/2}$ tail behavior seen in \cite{kovtunlec}
\begin{equation}
  \label{t0xdiv}
 \delta T^{0x} \propto \int \frac{d^3 \vec{k}}{(2 \pi)^3} e^{ i \vec{k}
\vec{x}} e^{- |t|/\chi - \frac{\chi k^2}{2} |t| } \sim \frac{T^2}{( 4
\pi)^{3/2} } \Bigg[ \frac{\chi^{5/2}}{ |t|^{3/2} } - \frac{\chi^{3/2}}{
( |t| )^{1/2} } + \dots \bigg]
\end{equation}
which shows the tail described in \cite{kovtun}.   Note that the short-time divergence $t\rightarrow 0$ generally does not commute with the limit of lack of polarization, $\chi \rightarrow 0$.  The introduction of the relaxation time, in the next section \ref{reltimeg} will cure this divergence.

\section{Relaxation time \label{reltimeg}}
As shown in \cite{gt2,gt3},  the polarization responding immediately to small external field according to Eq. \ref{parall} generally violates causality.
As consequence, polarization cannot appear immediately when a vorticity field is turned on. We have to specify a "minimal" time delay, in analogy with the Maxwell-Cattaneo equation \cite{rombook,kodama}.   

Adjusting the previous formalism to this realization, the Hermitian operator has to decay in absence of a vorticity field with a characteristic time scale $\tau_Y$.   Equation \ref{IS}, derived in \cite{gt3} would then appear by linking the expectation value of the magnon to a classical vorticity source.

Differentiating this equation in relation to time, we can write the equation of motion. According to conservation of total angular momentum at thermodynamic equilibrium, $\partial_\lambda S^{\lambda, \mu \nu} = - 2 T^{\mu\nu}_A$, where $S^{\lambda, \mu \nu}$ is the spin tensor and $T^{\mu\nu}_A$ the canonical antisymmetric energy-momentum tensor \cite{brauner}, and a partial integration substitution from Laplace transform. We get

\begin{equation}
\label{yeom}
  \langle Y^{\mu\nu}( \vec{k}, z) \rangle = \left(iz + \tau_Y z^2\right)^{-1}\left(- (1-iz \tau_Y) \chi(b_0,0) \omega^{\mu\nu}(\vec{k},0) - 4 w_0 \chi^2(b_0,0) g^{\rho \{ \mu  }P^{ \beta \} \gamma }_\nu \underbrace{\int^{\infty}_0 dt e^{izt} \partial_\beta {\omega^{\rho \gamma }(\vec{k},t)}}_{\rightarrow (1-\delta_{\beta 0})(k \times \vec{\tilde{\pi}})_{\beta}^{\rho \gamma} /(iz) } \right) 
\end{equation}
the underbrace defines the linearized limit where $\vec{\tilde{\pi}}$ is perpendicular to the vorticity plane defined by $\rho \gamma$, assuming the initial condition with no polarization
$\langle Y^{\mu\nu}(0 , |\vec{k}| )\rangle = \chi \omega^{\mu\nu}(k)  \qquad \frac{\partial  }{\partial t} \langle Y^{\mu\nu } \rangle|_{t=0} = 0   $

Given the dynamics of equations \ref{IS} and \ref{yeom}
we redefine in Fourier space polarization to an asymptotic state to which the fluid relaxes.  In the linear approximation, we can substitute $Y^{\mu \nu}$ by a correction on $\chi$.
\begin{equation}
\label{relaxdef}
  Y^{\mu\nu} = \frac{y^{\mu\nu}}{1 + i \omega \tau_Y } \Longrightarrow \chi \rightarrow \frac{\chi}{1 + i \omega \tau_Y} 
\end{equation}

To appreciate the power of this substitution, we should recall the hydrodynamics poles from polarizeable fluid suffer from unphysical behavior as well as unstable one  \cite{gt3} and relaxation (unlike Maxwell-Cattaneo, it is first order) is needed to stabilize the behavior. Let us focus in the transverse green function and apply the above substitution

\begin{equation}\begin{aligned}
G^{\,R}_{ \, T^{xy} \, T^{xy} }(w,k) &= 4 i B \omega^3 \, \,  \longrightarrow  \, \,  \, \,  \frac{ i w_0 \chi^2  \omega^3}{(1 + i \omega \tau_Y)^2}
\end{aligned}\end{equation}

This correction turns the group velocity modes bounded and provides a stable solution. For instance, after enforcing causality the poles of eq. \ref{example} corresponds to the same evaluated in \cite{gt3}. The transport coefficient from the imaginary part of retarded Green Function of eq. \ref{example}, we have

\begin{equation}
\lim_{\omega \to 0}  \frac{1}{\omega^3} Im G^R_{T^{xy},T^{xy}} (\omega, \vec{k}) = w_0 \chi^2 , \quad \lim_{\omega \to 0} \lim_{k_z \to 0} -\frac{\omega}{k^2} Im G^R_{T^{00},T^{00}} (\omega, \vec{k}) = w_0     
\end{equation}

\begin{equation}
\lim_{\omega \to 0} \frac{\partial_\omega}{3}  \lim_{k_z \to 0} - \frac{\partial^2_k}{2} G^{\,R}_{ \, J^{0} \, T^{00}}(w,k) = w_0 \chi^2, \quad
\lim_{\omega \to 0} \lim_{k_z \to 0} - \frac{\omega^2}{2} Im G^{\,R}_{J^z \! \omega^{0z}}(w,k) = \frac{w_0\chi^2}{\tau^2_Y}      
\end{equation}
as discussed in \cite{kaja}, $ \lim_{\omega \to 0} \lim_{k \to 0} \neq \lim_{k \to 0} \lim_{\omega \to 0}$. The integral representation of retarded polarization-polarization correlator introduces variational of energy-momentum in response from metric or vorticity field perturbation $S^{\gamma \lambda}$.

\begin{equation} 
\delta T_{\mu\nu} (t, x) =  \int d \omega d^3 k e^{-i \omega t + i \vec{k}\cdot \vec{x}} G_{\mu\nu, \gamma \lambda} (\omega, \vec{k}) S^{\gamma \lambda} 
\end{equation}

The structure of Green function allows us to extract singular modes whose dispersion relation determine the behavior of system in the hydrodynamic limit $\omega,k \ll T$ for $G_{00,00}$ and $G_{01,01}$

The transverse perturbation is


\begin{equation}
   \label{longpertchi}
 \delta T^{0x} \propto \int  \frac{d^3 \vec{k}}{(2 \pi)^3} e^{ i \vec{k} \vec{x}} e^{- |t|/\chi  -  \frac{k^2 |t|}{2 \chi} ( \tau^2_Y + \chi^2) + i \tau k^2   }  \sim
 \frac{T^2}{ (4 \pi)^{3/2} } \Bigg[ \frac{\chi^{5/2}}{(( \frac{\tau_Y^2}{\chi^2} + 1  )|t|)^{3/2} } -  \frac{\chi^{3/2}}{ 2 ( \frac{\tau_Y^2}{\chi^2} + 1  )^{3/2} ( |t| )^{1/2} } + \dots   \bigg] 
\end{equation}
The causality bound derived in \cite{gt3} implies
$\frac{\tau_Y^2}{\chi^2} > 1$, so when $\chi \rightarrow 0$ Eq \ref{longpertchi} goes to zero as required.

Since we are most likely dealing with a non-renormalizeable theory, where the cutoff is physical, we factorize the exponential $e^{- |t|/\chi}$ in Eq. \ref{longpertchi} up to first order $1/\chi$.  This exponential play an important role of mass in the dispersion relation.  Higher order terms should be dropped in the infrared limit because they are irrelevant. The next leading term of the integral with the expansion of $cos (\textbf{k}^2 \tau_Y t ) $ above will be $\sim \frac{T^2 \tau_Y \chi^{2/3}}{(\frac{\tau^2}{\chi^2} + 1 )^{5/2} |t|^{3/2} }$.  

Note that the dependence of the correlator on time, with the long time tail is the same as was calculated in \cite{kovtun}, using similar methods  but applied  to thermodynamic fluctuations.    This is not so surprising:  As noted earlier
\cite{gt1} fluctuations and polarization depend on the same dimension of operators, and hence it is natural to expect their correlation to scale similarly.  The difference is that in this case we are correcting a non-dissipative theory, rather than an already-dissipative one as in \cite{kovtun}.

\section{Interactions of hydrodynamic modes \label{interhyd}}
\subsection{One loop effective action \label{seclint}}
Now we try to compute backreaction corrections to $\lnz$, introduced in section \ref{corrfunc} using perturbative techniques.  One can do this by means of an effective field theory one, where the symmetries and light degrees of freedom are the essential ingredients to analyze the form of the effective lagrangian.
As a starting point, we examine the structure of the full action setting by slow sound perturbations ($\phi^I$) and fast microscopic ($y^{\mu\nu}$) degrees of freedom. Here, we  consider a weak interaction between them. So the ``bare'' local action reads

\begin{equation}
S[\phi^I , y^{\mu\nu} \, ] \simeq \int d^4 x \left[  \lag_0 [ \phi^I \, ] + \lag_{y} [y^{\mu\nu}\,] + \lag_{\rm int } [\phi^I , y^{\mu\nu} \,] \; 
 \right] \end{equation}

The $\lag_0 [\phi^I]$ encodes, alone, the general idea of "standard" hydrodynamic, while $\lag_{y} [y^{\mu\nu} \, ]  $ regulates the dynamics of polarization variable.   Now, we will concentrate our efforts to work out on $\lag_{\rm int }$ sector. Conventionally, the couplings between light and heavy should be treated as small in order to not break down the perturbative expansion. 
Let us now  integrate out the fast $y$ sector

\begin{equation}
  \label{tracingout}
\lnz  = \ln \left[    \int \mathcal{D} y^{\mu\nu} \mathcal{D} \phi_I e^{S[\phi^I,y^{\mu\nu}]} \right] \rightarrow \ln \left[ \int \mathcal{D} \pi_L \mathcal{D} \pi_T \exp \left[ i \int d^3 x d t   \left(\lag_{o} + \lag_{int} + \lag_{self-int} \right)  \right] \right]
 \end{equation}
where and $\pi_{L,T}$ will be defined as  \cite{gt1,gripaios}.   Magnon fields will be incorporated with existing degrees of freedom via the prescription of section \ref{reltimeg}. 

 We expand the action up to fourth order, using Eq. \ref{staticdef} and following \cite{gt2}, for linear terms of up to 2 gradients we have


\begin{eqnarray}\label{fdsa}
\lnz_0
\simeq  1+
   A \left\{ [\partial\pi]  -\frac{1}{2}[\partial\pi^T \cdot\partial\pi] -\frac{1}{2} \dot\pi^2 \right\} + 
 \left(\frac{1}{2}A + C\right)[\partial\pi]^2.
\end{eqnarray}
where the constants $A,C$ are
\begin{equation}
c_s^2 = \underbrace{\frac{2F''(b)}{F'(b)}}_{<0}+1,\qquad A = T_0 F^\prime(b_0), \qquad C = \frac{1}{2}b_0^2 F^{\prime\prime}(b_0),
\end{equation}

where the enthalpy is $w_0 = - F_b(b_0)$, $b_0 = 1$ and has the dimension $[M^4]$. Note that here $\partial \pi^T$ means the transpose of the matrix of $\pi$s rather than the transverse direction. The first term of Lagrangian Eq. \ref{fdsa} indicates the minimized potential by $w_0$, the second one vanishes by integration. The third and fourth are the well known free propagation of sound-waves and the fifth transverse (strong coupling) excitations of $\phi^I$.  Of course there are vortex-sound interactions, however, they must be neglected if we compared with other terms of Lagrangian $\sim \chi^2$.

Inverting Eq. \ref{fdsa} will give the propagator for the sound-waves in momentum space $G_{\partial \pi_i,\partial \pi_j}$ \cite{gripaios} ($\partial \pi$ represents the general matrix $\partial_i \pi_j$, including both $\pi_T$ and $\pi_L$, $[\pi]$ is the trace, defining the sound-wave
\begin{equation}
  \label{proppi}
  G_{[\partial \pi][\partial \pi]}=\frac{ik^2}{w^2-c_s^2 k^2} \eqcomma G_{\dot{\pi}_l[\partial \pi]}=\frac{i w k_l}{w^2 - c_s^2 k^2} \eqcomma G_{\dot{\pi}_l \dot{\pi}_m  }= i\delta_{lm} +\frac{i c_s^2 k_l k_m}{w^2 -c_s^2 k^2}
  \end{equation}
It can be seen by inspection that the infrared limit $w \rightarrow 0$ of transverse mode propagators diverges, confirming the potential instability of \cite{nicolis1}.

The interaction term between cell fluid and polarization $\sim F (-b y^2) $.

\begin{eqnarray}
\lnz_{int} &\simeq&  w_0 \chi^2 (b_0,0) \Big\{  \bigg([ \partial \pi]  [\partial \dot \pi \cdot \partial \dot \pi ] + [ \partial \pi]  (\partial_\mu \dot \pi) \cdot (\partial^\mu \dot \pi) \bigg) (  1 +  c_s^2 + \underbrace{ \frac{2 \chi_b}{\chi}}_{self-int})  \nonumber \\ \nonumber \\ 
 && \quad \left. \vphantom{\frac{}{}}  - \left[  \sfrac12 \dot {\vec \pi}^2  + \sfrac12 [\di \pi^T \di \pi ] \right] \left[ (\partial_\mu \dot \pi) \cdot (\partial^\mu \dot \pi) +  [\partial \dot \pi \cdot \partial \dot \pi ] \right] ( 1 +  \frac{f^{\prime \prime}}{ 2 f^{\prime}} + \underbrace{ \frac{2 \chi_b}{\chi}}_{self-int} ) \right. \nonumber \\ \nonumber \\ 
&& \quad [\partial \pi]^2 \bigg( (\partial_\mu \dot \pi) \cdot (\partial^\mu \dot \pi) + [\partial \dot \pi \cdot \partial \dot \pi ] \bigg) (  \frac{f^{\prime \prime \prime}}{2 f^{\prime}} +  \frac{ 2 \chi_b}{\chi} c_s^2 +  \frac{c_s^2}{2} +  c_s^2 +  \frac{2 \chi_b}{\chi} +  1 + \underbrace{   \frac{2 \chi_b}{\chi} +  \frac{\chi^2_b}{\chi^2} +  \frac{\chi_{b b} }{\chi}}_{self-int} ) \nonumber \\ \nonumber \\
&& \quad 
\bigg(  2 \dot \pi \cdot \partial \ddot \pi \cdot \dot \pi   + [\partial \dot \pi \cdot \partial \dot \pi \cdot \partial \pi ] + ( \partial_\mu \dot \pi) \cdot ( \partial_\mu \dot \pi \cdot \partial \pi )  - \dot \pi \cdot \partial ( \partial^I \pi^J ) \partial^J \dot \pi^I - \dot \pi \cdot \partial ( \partial_\mu \pi^I ) \partial_\mu \dot \pi^I  \nonumber \\ \nonumber \\ 
&& \quad   + \ddot \pi \cdot \ddot \pi \cdot \partial \pi + 2 \ddot \pi  \cdot \dot \pi \cdot \partial \dot \pi + 2 \dot \pi \cdot \partial \dot \pi \cdot \ddot \pi \bigg) ( \underbrace{1}_{self-int} + [ \partial \pi ] c_s^2 + [\partial \pi] + \underbrace{  \frac{2 \chi_b}{\chi} }_{self-int} )
  \label{lintdef}
\end{eqnarray}
In the following paragraphs we shall link each line to Feynman diagrams.  The ``gluon'' (spring-like) lines will  be used for longitudinal modes, and the ``photon'' (wavy) lines for transverse ones (respectively sound and vortices when $\chi \rightarrow 0$).   In the limit where polarization vanishes the former correspond to vortices and the latter to sound waves.   Note that the Feynman diagrammatic structure does not change when the relaxation time $\tau_Y$ is present.  One must just update the vertex and propagator terms to those of Eq. \ref{relaxdef}, $\chi \rightarrow \chi \left(1 + w\tau_Y  \right)^{-1}$, following section \ref{reltimeg}, as Maxwell-Cattaneo is a constitutive equation,  not entering in the bare Lagrangian but, through the dissipation-fluctuation therem, the correlation function and hence $\lnz_{eff}$.

The perturbation expansion breaks down when the energy $E^2 \sim \chi^{-2} $ where the dimension of $[\chi] = M^{-1}$ and $\pi^I = M^{-1}$. We will analyze the leading terms in according with $E \ll \chi^{-1}$. The strength interaction for a typical tree level process is $\mathcal{M}_{TT \rightarrow TT} \sim \frac{p^6}{w_0 \chi^{-2}}$.

\begin{figure}
     \epsfig{width=0.39\textwidth,figure=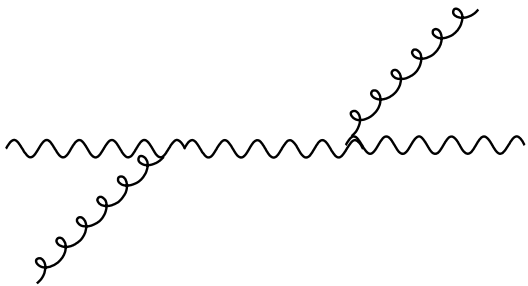}
     \caption{\label{soundvort1} Sound-vortex scattering
       }
\end{figure}
\begin{figure}
\epsfig{width=0.49\textwidth,figure=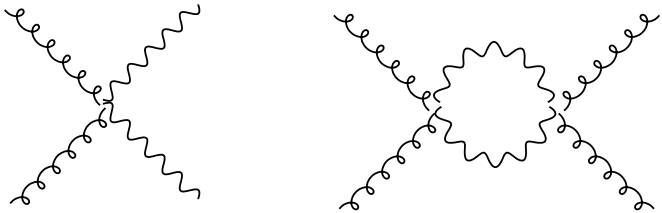}
\caption{\label{soundloop1} Production of vortices by sound-waves (left) and scattering of sound waves by an intermediate vortex state}
\end{figure}

The Feynman rules are as follows: each vertex a $w_0 \chi^2$, $1/w_0$ internal line, and $1/\sqrt{w_0} $ external line. We extract fourth order terms whose tree level diagram has one legs as longitudinal perturbation propagation $ \sim [\partial \pi] \mathcal{O}(\pi^3)$ by symmetry arguments, and terms without spatial derivative must vanish since they do not contribute on vortex amplitude but for longitudinal one.

The representation of the first line of Eq. \ref{lintdef} corresponds to the Feynman Diagram in Fig \ref{soundvort1}.
The vertical and horizontal lines are space and time arrow. The process is similar to Rayleigh scattering in Electrodynamics where a non-linear physical object (vortex) absorbs and emits one or more sound waves by a non-linear harmonic process. Only unsteady vortex rather than stationary one perpendicular to scattering plane provide variation of longitudinal velocity, so that  vortex can participate of this process where the emitted sound wave has the same frequency of the incident one.

The second and third line of Eq. \ref{lintdef} in Feynman diagram language are shown in Fig. \ref{soundloop1}. 
They represent a process studied non-relativistically in \cite{westerveldt}, the production of vortices by soundwaves and the associated sound wave scattering. The first diagram indicates the cross section between sound waves and vorticity distribution of a turbulent flow. The vertical and horizontal lines determine the space and time arrow. The interaction of vortex and sound describes convective propagation of a weak vorticity fluctuation.
In \cite{gtstart}, it was shown that the absence of the energy gap and propagation velocity of \cite{nicolis1} lead to a badly divergent loop.   As is shown here, as hypotesized in \cite{gt1,gt3} the divergence is mitigated by polarization, as the ratios with $\chi$ as denominator cut off the divergences.

\begin{figure}
     \epsfig{width=0.39\textwidth,figure=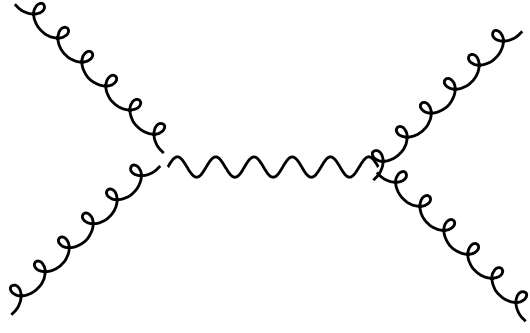}
     \caption{\label{soundtree1} Vortex formation by soundwaves
       }
\end{figure}
Note that there are some variations at the orientations of these diagrams as t,u - channel due to the lack of Lorentz covariance in the $IJ$ indices, see \cite{nicolis1} for a discussion on this.

The sound vorticity coupling (Fig. \ref{soundtree1}) arises when non-linear perturbative hydrodynamical variables are taken into account in the fluid equations of motion, and has the potential to produce both vortices out of sound-waves and vice-versa (when the diagram is split). Looking at the third line of   Eq.\ref{lintdef} it is hoever clear that quantitatively usually vortex production dominates over sound production.

Finally, the coupling affects convection, expansion and stretching of vorticity configuration: The second-to-last and last lines of Eq.\ref{lintdef} describe the interaction of third order in $\pi$ and gives the leading radiative order diagrams

\begin{figure}
  \epsfig{width=0.59\textwidth,figure=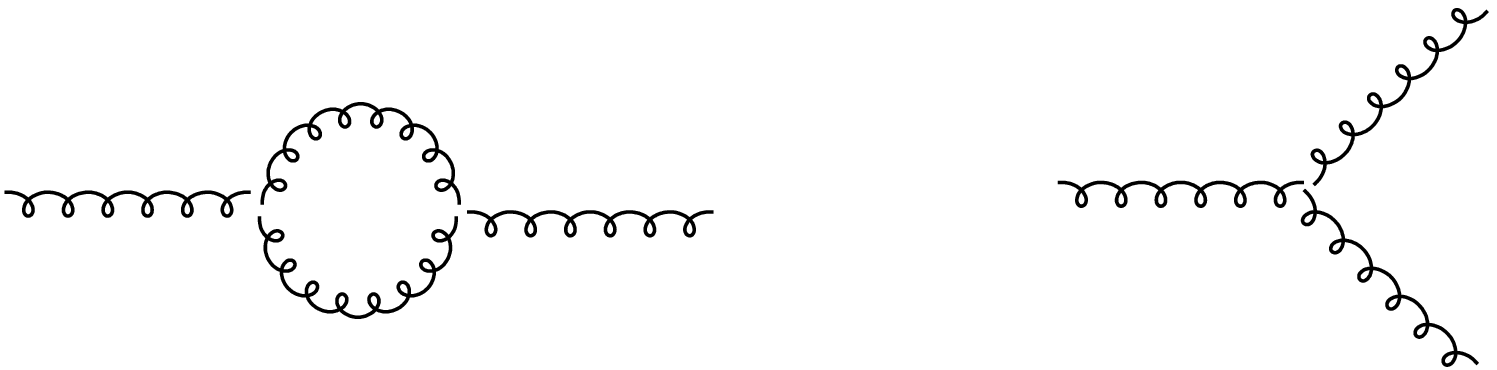}
  \epsfig{width=0.59\textwidth,figure=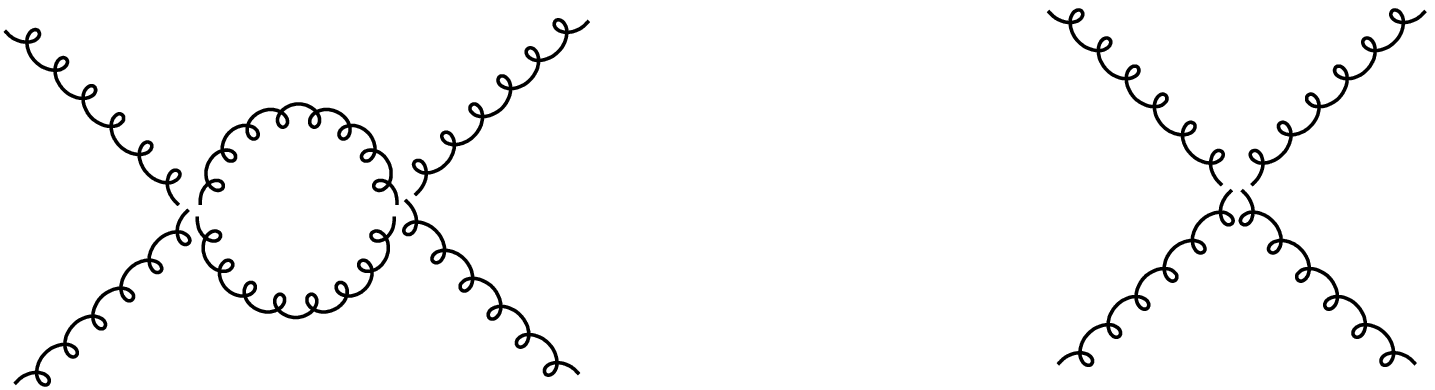}
     \caption{\label{vortloop1} Sound-sound backscattering and scattering
       }
\end{figure}
The Vorticity-Vorticity interaction has been analyzed by \cite{lighthill}, as has the generation of sound by vorticose sources.
In a relativistic setting this interaction produces infrared divergences, as the vortex has no energy gap and no propagation speed \cite{nicolis1}.   Once we include, and integrate out,  spin-spin interaction,  the respective orders of spin-spin interaction in the effective theory become $\sim \sqrt{w_0}^{-1} \chi^2$ and $ \sim w_0^{-1} \chi^4$

Equation \ref{tracingout} shows that,
without loss of generality, we can take upon the Lagrangian Eq. \ref{eos} only the spin interactions with the background $b_0$, in other words, $\sim F ( - b_0 y^2 (\pi) )$. Thus the spin interacts with the vacuum of fluid (hydrostatic configuration) when the physical coordinate are aligned with the comoving one at given pressure.  After turning on the external field the vacuum's original symmetry of $SO(3)$ is broken to $SO(2)$. The $\pi^x$ and $\pi^y$ eat two Goldstone boson and become massive, whereas the $\pi^z$ remains massless. The Lorentz boost is broken. The self-interacting lagrangian up to fourth order is (Fig. \ref{vortloop1})

\begin{equation}\begin{aligned}
& \mathcal{L}_{self-int} \sim  w_0 \chi^2 (b_0,0) \bigg\{ \bigg[  (\partial_\mu \dot \pi) \cdot (\partial^\mu \dot \pi) + [\partial \dot \pi \cdot \partial \dot \pi ] \bigg] +
\\ \\ 
&  \dot \pi \cdot \partial \pi \cdot \partial \ddot \pi \cdot \dot \pi + \dot \pi \cdot \partial \pi \cdot \partial \dot \pi \cdot \ddot \pi + 2 \ddot \pi \cdot \partial \pi \cdot \partial \dot \pi \cdot \dot \pi + \dot \pi \cdot \partial \dot \pi \cdot \partial \pi \cdot \ddot \pi + 2 (\ddot \pi \cdot \partial \pi^I) (\dot \pi \cdot \partial \dot \pi^I) - \ddot \pi \cdot \dot \pi \cdot \partial \pi \cdot \dot \pi 
\\ \\ 
& + 2 \dot \pi \cdot \partial (\partial^J \pi^I ) \dot \pi^J \ddot \pi^I +  [ \dot \pi \cdot \partial ( \partial^J \pi^I )] [\dot \pi \cdot \partial (\partial^I \pi^J)] - [ \dot \pi \cdot \partial ( \partial^J \pi^I )] [ \partial^I \dot \pi \cdot \partial \pi^J ] + (\partial_\mu \dot \pi \cdot \partial \pi ) \cdot (\partial_\mu \dot \pi \cdot \partial \pi )  \\ \\ 
&  + [\partial \dot \pi \cdot \partial \pi \cdot \partial \dot \pi \cdot \partial \pi] + [ \partial^\mu \dot \pi \cdot \partial \pi^I ] [ \dot \pi \cdot \partial (\partial^\mu \pi^I)] - (\partial^\mu \dot \pi \cdot \partial \pi ) \cdot ( \dot \pi \cdot \partial ( \partial_\mu \pi )) \bigg\}  \\ \\  
& + \chi^4 (b_0,0) \bigg\{ \left[ 2 (\partial_\mu \dot \pi ) (\partial_\mu \dot \pi ) [\partial \dot \pi \cdot \partial \dot \pi ]  +  (\partial_\mu \dot \pi )^2 \cdot ( \partial_\mu \dot \pi )^2 + [\partial \dot \pi \cdot \partial \dot \pi ]^2 \right] (  \frac{c_s^2}{2} +  \frac{\partial_{\omega^2} \chi(b_0,0)}{ \chi^3 } ) \bigg\} \\ \\ 
\end{aligned}\end{equation}
.  These results will be cemented by the explicity calculation of the propagator in the next subsection.

While this theory is non-renormalizeable and non-Lorentz-covariant, and it is spacetime diffeomorphism that are broken, the above process (some directions of the Goldsone boson are ``eaten'' via interaction with the consensate) has a similarity to the Higgs mechanism \cite{higgs}.   Instead of internal Gauge $SU(2)$ the condensate breaks spacetime ``Gauge'' $SO(3)$, and, as we will see in the next section, the broken generators correspond to the  ``Goldstone'' components ``eating'' vortex polarizations that aquire a mass gap, although, because of relaxation, it becomes a ``dissipative gap'' similar to \cite{massdiss}.

It is useful to consider diagrams for the following correlator at $\mathcal{O} (w_0^{-2} \chi^4 (b_0,0))$ and $\mathcal{O} (w_0^{-2} \chi^2 (b_0,0))$ to self-interaction and vortex and ''fluid'' interaction, respectively. 


\begin{equation}
\langle \, y^{ij} \, (x_1,t_1) , \, y^{pq} \, (x_2,t_2) \rangle = \langle \, \partial^i \dot \pi^j \, , \, \partial^p \dot \pi^q \, \rangle - \langle \, \partial^i \pi^j  \, , \, \partial^p \dot \pi \cdot \partial \pi^q \, \rangle + \dots  
\label{yleff}
\end{equation}
\begin{equation}
  \label{buy}
  \langle \, b u^0  \, (x_1,t_1) , \,  y^{ij} \, (x_2,t_2) \rangle = \ave{ y^{ij} \, (x_2,t_2)}+\langle \, [\partial \pi] \, , \, \partial^i \pi^j \, \rangle +  \frac{1}{2} \langle \, [\partial \pi^2 ] \, , \, \dot \pi^l \partial_j \partial_l \pi^i \, \rangle +  \dots
  \end{equation}
This is a diagrammatic way to understand how the divergence in Eq. \ref{proppi} is regularized by polarization.

\begin{figure}
     \epsfig{width=0.99\textwidth,figure=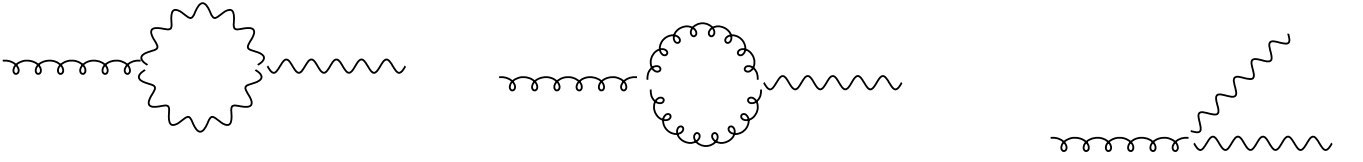}
     \caption{\label{sounddiags1} Sound-vortex conversion
       }
\end{figure}

Via the terms of $\mathcal{O} (w_0^{-2} \chi^2 (b_0,0))$, Fig \ref{sounddiags1} converts sound waves into vortices, and is also peculiar to mediums with a spin.   Physically, what happens is that the change in temperature during the compression of a sound perturbation changes $\chi$, which in turn modifies the vorticity. The reversed process, where a vortex perturbation changes the compressibility and releases sound is of course also there.  Hence, non-propagating vorticity fluctuations are now associated with an energy change of the background that can be used to emit sound and dissipate.  This resolves the singularity in the $S$-matrix definition pointed out in \cite{nicolis1}.

The extra gradient terms, when included in the propagator definition Eq. \ref{proppi}, will lead to higher powers of $w$ in the numerator which will cancel the divergence.   Equation \ref{hgf} in the next section does this explicitly.

We close by noting that 
the pressure generation by small vorticity fluctuation is irrelevant since the former depends upon the later by a higher order at $v(t,\vec{x})$, by EFT language $\sim c_s^2 F_{bb} [\partial \pi]^2 $ field fluctuation. 

In the next subsection, we will use the diagrammatic insights obtained here together with the results of section \ref{corrfunc} to compute the interaction part of the propagator explicitly.


\subsection{Propagator correction due to hydrodynamic interactions \label{backreac}}
In this subsection, we use our intuition from the previous section and the calculations in section \ref{corrfunc} to extend the dissipative phenomena investigated to backreaction from interactions {\em between fluid modes}. In doing so, we examine the effect on transport coefficient of hydrodynamic fluctuation-generated collective excitations. 
We start from the linearization of Eq. \ref{tmununotsym} and \ref{jdef}

\begin{equation}\begin{aligned}
& T_{\mu\nu} \sim \, 2 \, F^{\prime} (b_0) \, \chi^2(b_0,0) \, \bigg(  g_{I[\rho} \, \partial_{0]} \, \dot \pi^I \ddot \pi^\rho  + \, \dot \pi^I H_{IJ} \, \dot \pi^J \bigg) \delta^{\mu\nu} +  2 F^{\prime} (b_0) \chi (b_0,0) y_{\rho \nu} \delta^\mu_0 + \frac{1}{3} F^{\prime} (b_0) y_{\rho \sigma} \epsilon^{\mu \rho \alpha}\epsilon_{0 \nu J} \partial^\sigma \partial^\alpha \pi^J + \dots \\
& J^\mu_I \, \sim \,  \,  F^{\prime} (b_0) \, \chi(b_0,0) \, \partial_0 \omega^{\mu}_{I} \, - 2 F^{\prime} (b_0) \, (  y^2 \, - \, b_0 \, \chi \, \partial_b \chi \, \omega^2 ) \, \delta^\mu_I \, \dots
\end{aligned}\end{equation}

where we used the projector
\begin{equation}
\label{hdefeq}
  H^{IJ}_{KL} = ( \delta^{IJ} \delta_{KL} + \delta^I_K \delta^J_L )
\end{equation}
the second order contribution to the propagators, labeled by $G^{(2)}$ henceforward, are 

\begin{equation}
  G_{J_i J_k}^{(2)}(t,\vec{x}) = \frac{1}{ w^2_0 } \langle \pi_m (t,\vec{x}) \pi_n (t,\vec{x}) \pi_q (0) \pi_p (0) \rangle_{eq} \, = \frac{1}{ w^2_0 } \, \langle \pi_m (t,\vec{x}) \pi_q (0) \rangle_{eq} \langle \pi_n (t,\vec{x})  \pi_p (0) \rangle_{eq} ,
\label{gij1}
\end{equation}
  \begin{equation}
    G_{T_{ij} T_{kl}}^{(2)}(t,\vec{x}) = \, \frac{1}{ w^2_0  } H^{ij}_{mn} H^{kl}_{qp} \langle \pi_m (t,\vec{x}) \pi_n (t,\vec{x}) \pi_q (0) \pi_p (0) \rangle_{eq} \, = H^{ij}_{mn} H^{kl}_{qp} \frac{1}{ w^2_0  } \, \langle \pi_m (t,\vec{x}) \pi_q (0) \rangle_{eq} \langle \pi_n (t,\vec{x})  \pi_p (0) \rangle_{eq}
    \label{gij2}
\end{equation}
With the structure of the coefficients determined via Eqs \ref{yleff} and \ref{buy}.
Note that $SO(3)$ symmetry keeps vortex-sound and sound-vortex coupling, $\langle \pi_L (x) \pi_T (0) \rangle$ and $ \langle \pi_T (x) \pi_L (0) \rangle$, equal.

Even though the correlation functions in Eq. \ref{gij1} and \ref{gij2} may appear similar, it is important to remember that they come from varying w.r.t. different generators.

By treating the fluctuations as Gaussian, we can order the operator in correlation function in such a way 
\begin{equation}
  \label{tijtijexp}
 G_{T_{ij} T_{kl}}^{R (2)}(\omega,\k) = \frac{2}{w_0^2}
H^{ij}_{mn} H^{kl}_{pq}\,
\int\! \frac{d\omega'}{2\pi} \frac{d^dk'}{(2\pi)^d}\, G^{\,R (1)}_{ \, T^{0m} \, T^{0p} }( \omega^{\prime}, k^{\prime} ) \,  G^{\,R (1)}_{ \, T^{0n} \, T^{0q} }( \omega^{\prime}, k^{\prime}) =  \dots + \mathcal{O}(M^4) + \dots 
\end{equation}

where $M$ is an ultraviolet cutoff and we use the approximate process of symmetrization of imaginary green function described in \cite{kovtunlec}. 

\begin{equation}\begin{aligned}\label{GT}
& G^{\,R}_{ \, T^{0i} \, T^{0j} }(\omega, \textbf{k}) = \frac{(w_0 T)}{2} \bigg[ \bigg( \delta^{ij} - \frac{k^i k^j}{\textbf{k}^2} \bigg) \frac{ 16 \chi^4 (\omega^4\textbf{k}^2 - \omega^2\textbf{k}^4) - 2  \chi^2 (\omega^4 -\textbf{k}^2 \omega^2) }{4 \chi^2 (\omega^4 - \textbf{k}^2 \omega^2) - \omega^2 }  \\   & + \bigg(\frac{k^i k^j}{\textbf{k}^2}\bigg) \frac{ \omega^2 -   \chi^2 (3 \omega^4 + 2 \omega^2 c_s^2\textbf{k}^2) +   \chi^4 (3 \omega^6 + 2 \omega^4 \textbf{k}^2)  }{ \chi^2 (\omega^4 + \textbf{k}^2 \omega^2) -  \omega^2 + c_s^2 \textbf{k}^2}  \bigg]
\end{aligned}\end{equation}
When interactions are turned on the SO(3) symmetry breaks down and the $\pi$ becomes massive. The transverse part becomes strongly dependent of $\chi$. If $\chi \rightarrow 0$, then it yields $G_{0x,0x} \rightarrow 0$. The transverse part may be then simplified 

\begin{equation}\label{hgf}
G^{\,R}_{ \, T^{0i} \, T^{0i} }(w,k) = - \frac{w_0}{2} \bigg( 1 - \frac{(k^i)^2}{\textbf{k}^2} \bigg) \frac{( (\omega^2 + k^2) \chi^2 k^2 + 1 ) (\omega^2 - k^2)}{(\omega^2 - k^2) -  \frac{1}{ \chi^2}  } 
\end{equation}

It is easy to see the denominator part of retarded green function above corresponds to $p^2 - m^2$, where the mass $m \sim (2 \chi)^{-1}$. We must remember that, because of the symmetries of fluid dynamics our green function will never look like as green function of ''free particle'' but rather as interacting green function, where the numerator is different from $i$.
As we have justified previously the propagator is, in the infrared limit, that of a massive vector particle, thereby realizing the effective vortex mass conjectured in \cite{landau}.

\begin{equation}\label{hgf}
G^{\,R}_{ \, T^{0i} \, T^{0i} } = - w_0 \bigg( 1 - \frac{(k^i)^2}{\textbf{k}^2} \bigg) \frac{ (\omega^2 - k^2) }{(\omega^2 - k^2) -  \frac{1}{\chi^2}  } 
\end{equation}
Note the similarity of this propagator to one of a massive vector particle, confirming the analogy with the Higgs mechanism argued for in the previous sub-section.

Eq. \ref{hgf} can be used in Eq. \ref{tijtijexp} to obtain the propagators of all the other components of the energy-momentum tensor.
Using dimensional regularization we obtain, in the $k=0$ frame $G_{T_{ij} T_{kl}}^{R (2)}(\omega,\k= 0)=$
\begin{equation}\begin{aligned}
 = \frac{ T^2 \chi^4 H^{ij}_{kl} }{(4\pi)^2} \mu^{2 \epsilon} \bigg[ \frac{1}{\hat{\epsilon}} (2 M^2 + M^2 p^2 + M^4) + \frac{2}{3} (\frac{p^4}{5} - M^2 p^2 -3 M^4 ) + \int^1_0 dx ( 3 a^4 + 6 a^2 x^2 p^2 + x^4 p^4) ln (a^2/\mu^2) \bigg]
\end{aligned}\end{equation}

where $a = p^2 x (x-1) + M^2$, $1/\hat{\epsilon} = 1/ \epsilon + \gamma_e + \mathcal{O}(\epsilon)$, $\mu^{2 \epsilon} = 1 + 2 \epsilon ln \mu $. After algebra  manipulations, the cutoff-independent part of the propagator reduces to

\begin{equation}
 G_{T_{ij} T_{kl}}^{R (2)}(w)= \frac{T^2 H^{ij}_{kl} }{(4\pi)^2} (-1 - \frac{2}{3} \chi^2 + ( 6 + 2 \omega^2 \chi^2 +  \frac{\chi^4}{4} )  ln(\frac{1}{\chi^2 \mu^2 }))
\end{equation}
which remains finite for any non-zero $\chi^2$ but diverges at $\chi \rightarrow 0$.

Thus we have explicitly confirmed the intuition of \cite{gt1} at the level of the propagator: once 1-loop corrections are included, the presence of vortical susceptibility, by giving a mass-like gap determined by $\chi$ to vortices, stabilizes the vorticity divergences described in \cite{nicolis1} 
\section{Discussion and conclusion}

We used some common effective field theory techniques to study the correlation function of hydrodynamic variables. In the section \ref{pqpqp}, the green functions arising from the linear response theory established an easy way to characterize the dynamics of a system with spin and vorticity close to equilibrium. Thus, we could relate transport coefficients of polarization $\chi$ (examined in section \ref{chisec}) and $\tau_y$ (examined in section \ref{reltimeg}) to correlators. 
In order to generalize to correlation functions of non-conserved quantities, in section \ref{corrfunc} we used variational approach where the background perturbation of the vortical field and the metric produce all possible fluctuations in according with the symmetry of the system.

These green functions contain information on how the presence of external vortex field modify the structure of our hydrodynamic with spin, which as we find can give an effective mass to the vortices. Under the presence of an external vortical source the symmetry under reversal time, homogeneity and isotropy are no longer valid. Initially, we have $h_{xx}= h_{yy} = h_{zz}$ due to $SO(3)$ group and after turning on the source, the pressure changes from the usual form of ideal fluid. Therefore, the direction along the axial rotation axes is $P_\perp =  \frac{1}{2} T^a_a =\frac{1}{2}(T^x_x + T^y_y)$, whereas the perpendicular direction corresponds to $P_\parallel = \frac{1}{2}T^z_z $. Each perturbation relates different process of such a system, the transverse direction contributes to aligning of spins ( polarization), while the longitudinal one corresponds to standard convective processes.  In section \ref{seclint} and \ref{backreac} we relate these to the one-loop effective action and higher order corrections of linear hydrodynamics.

Dissipation,  given by the magnon relaxation to equilibrium, has proven to be necessary to preserve causality.  Including it results in long time-tails of the correlators involving vorticity, in analogy to the correlators involving thermal fluctuations of the type studied in \cite{kovtun}.   This confirms the intuition, described in \cite{gt1}, that polarization from vorticity and hydrodynamic backreaction from thermal fluctuations arise at the same order in effective theory expansion.   Given the concurrent experimental discovery of hydrodynamic phenomena in small systems \cite{small} and polarization \cite{beclisa}, these relations give us hope to pin down quantitatively the effect of microscopic fluctuations on hydrodynamic evolution from experimental data.

An obvious extension is to include the microscopic shear and bulk viscosities, and study their interplay with the transport coefficients examined in this section.
This could be done within a Schwinger-Keldysh formalism \cite{floer,gt0,comer}, and will have to be left to a forthcoming work.

Another issue left for further work is the discussion of microscopic non-Abelian Gauge symmetry, which is in practice the interchange between angular momentum (carried by vorticity) and polarization (carried by $y_{\mu \nu}$ and $Y_{\mu \nu}$.   Mathematically, this would be achieved by making sure all correlation functions are gauge-covariant.   Earlier literature \cite{ghosts} showed that this cannot be achieved so easily, and indeed the left hand side of Eqs such as \ref{ymunueq} would have different transformation properties from the right hand side.  As \cite{ghosts} suggests, non-hydrodynamic modes might be necessary to resolve this ambiguity.

Finally, a connection with kinetic theory techniques, of the type used in  \cite{flork1,flork2,flork3,gale,dirk} will be necessary to provide a microscopic description of the coefficients described in this work.    As written in the introduction, this needs to be left for a forthcoming work, although progress in this area is rapid.   For example, the non-locality of the collision term derived in \cite{dirk} imposes limits on relaxation time comparable to the causality requirements discussed here.

In conclusion, in this work we investigated the linear response and fluctuation-dissipation properties of the theory developed in \cite{gt1,gt2,gt3}, We found that these early results are consistent
with linear response theory, derived fluctuation-dissipation relations
and built an effective lagrangian to one-loop, confirming that
microscopic polarization can act as a ``Higgs mechanism'' for vorticity,
giving a mass to vortices and stabilizing the theory. This confirms the
intuition of \cite{landau}, and could lead to a stable theory which includes microscopic fluctuations.
We also found that the  long-time tales in the correlators between vortical variables behave analogously to the fluctuation tails studied in \cite{kovtun}, confirming the relation between thermodynamic fluctuations and polarization.
Phenomenological applications of this theory to heavy ion collisions and cosmology can now be developed.

GT acknowledges support from FAPESP proc. 2017/06508-7,
partecipation in FAPESP tematico 2017/05685-2 and CNPQ bolsa de
produtividade 301432/2017-1. DM thanks CAPES for support This work is a part
of the project INCT-FNA Proc. No. 464898/2014-5.


\begin{thebibliography}{}

  

   \bibitem{gt1}
  D.~Montenegro, L.~Tinti and G.~Torrieri,
  Phys.\ Rev.\ D {\bf 96}, no. 5, 056012 (2017)
  Addendum: [Phys.\ Rev.\ D {\bf 96}, no. 7, 079901 (2017)]
  doi:10.1103/PhysRevD.96.079901, 10.1103/PhysRevD.96.056012
  [arXiv:1701.08263 [hep-th]].

\bibitem{gt2}
  D.~Montenegro, L.~Tinti and G.~Torrieri,
  Phys.\ Rev.\ D {\bf 96}, no. 7, 076016 (2017)
  doi:10.1103/PhysRevD.96.076016

  
\bibitem{gt3} 
  D.~Montenegro and G.~Torrieri,
  Phys.\ Rev.\ D {\bf 100}, no. 5, 056011 (2019)
  doi:10.1103/PhysRevD.100.056011
  [arXiv:1807.02796 [hep-th]].

\bibitem{flork1} 
  W.~Florkowski, B.~Friman, A.~Jaiswal and E.~Speranza,
  Phys.\ Rev.\ C {\bf 97}, no. 4, 041901 (2018)
  doi:10.1103/PhysRevC.97.041901
  [arXiv:1705.00587 [nucl-th]].

\bibitem{flork2} 
  W.~Florkowski, R.~Ryblewski and A.~Kumar,
  Prog.\ Part.\ Nucl.\ Phys.\  {\bf 108}, 103709 (2019)
  doi:10.1016/j.ppnp.2019.07.001
  [arXiv:1811.04409 [nucl-th]].

  \bibitem{flork3}
S.~Bhadury, W.~Florkowski, A.~Jaiswal, A.~Kumar and R.~Ryblewski,
[arXiv:2002.03937 [hep-ph]].

\bibitem{gale} 
  S.~Shi, C.~Gale and S.~Jeon,
  arXiv:2002.01911 [nucl-th].

\bibitem{hongo} 
  K.~Hattori, M.~Hongo, X.~G.~Huang, M.~Matsuo and H.~Taya,
  Phys.\ Lett.\ B {\bf 795}, 100 (2019)
  doi:10.1016/j.physletb.2019.05.040
  [arXiv:1901.06615 [hep-th]].

\bibitem{bec}
  F.~Becattini, L.~Bucciantini, E.~Grossi and L.~Tinti,
  Eur.\ Phys.\ J.\ C {\bf 75}, no. 5, 191 (2015)
  doi:10.1140/epjc/s10052-015-3384-y
  [arXiv:1403.6265 [hep-th]].

  \bibitem{nair}
D.~Karabali and V.~Nair,
Phys. Rev. D \textbf{90}, no.10, 105018 (2014)
doi:10.1103/PhysRevD.90.105018
[arXiv:1406.1551 [hep-th]].

\bibitem{dirk}
N.~Weickgenannt, E.~Speranza, X.~l.~Sheng, Q.~Wang and D.~H.~Rischke,
[arXiv:2005.01506 [hep-ph]].

    \bibitem{nicolis}
  S.~Dubovsky, L.~Hui, A.~Nicolis and D.~T.~Son,
  arXiv:1107.0731 [hep-th].

  
  \bibitem{nicolis1}
  S.~Endlich, A.~Nicolis, R.~Rattazzi and J.~Wang,
  JHEP {\bf 1104}, 102 (2011)
  [arXiv:1011.6396 [hep-th]].

  
\bibitem{gt0} 
  D.~Montenegro and G.~Torrieri,
  Phys.\ Rev.\ D {\bf 94}, no. 6, 065042 (2016)
  doi:10.1103/PhysRevD.94.065042
  [arXiv:1604.05291 [hep-th]].

  \bibitem{groz}
  S.~Grozdanov and J.~Polonyi,
  Phys.\ Rev.\ D {\bf 91}, no. 10, 105031 (2015)
  doi:10.1103/PhysRevD.91.105031
  [arXiv:1305.3670 [hep-th]].



  \bibitem{floer} 
  S.~Floerchinger,
  JHEP {\bf 1609}, 099 (2016)
  doi:10.1007/JHEP09(2016)099
  [arXiv:1603.07148 [hep-th]].

  \bibitem{jackiw}
R.~Jackiw, V.~Nair, S.~Pi and A.~Polychronakos,
J. Phys. A \textbf{37}, R327-R432 (2004)
doi:10.1088/0305-4470/37/42/R01
[arXiv:hep-ph/0407101 [hep-ph]].
  
\bibitem{comer}
N.~Andersson and G.~Comer,
Class. Quant. Grav. \textbf{32}, no.7, 075008 (2015)
doi:10.1088/0264-9381/32/7/075008
[arXiv:1306.3345 [gr-qc]].
  

  \bibitem{beclisa}
F.~Becattini and M.~A.~Lisa,
doi:10.1146/annurev-nucl-021920-095245
[arXiv:2003.03640 [nucl-ex]].


\bibitem{becwigner} 
  F.~Becattini,
  arXiv:2004.04050 [hep-th].
 

\bibitem{rombook} 
  P.~Romatschke and U.~Romatschke,
  doi:10.1017/9781108651998
  arXiv:1712.05815 [nucl-th].

  
\bibitem{kodama}
R.~Derradi de Souza, T.~Koide and T.~Kodama,
Prog. Part. Nucl. Phys. \textbf{86} (2016), 35-85
doi:10.1016/j.ppnp.2015.09.002
[arXiv:1506.03863 [nucl-th]].

  
  
    \bibitem{massdiss}
K.~Trachenko,
Sci. Rep. \textbf{9}, no.1, 6766 (2019)
doi:10.1038/s41598-019-43273-9
[arXiv:1905.07405 [cond-mat.stat-mech]].


  \bibitem{landau}
  L. D. Landau, Phys. Rev. {\bf 60}, 356 (1941).

  

\bibitem{gripaios}
  B.~Gripaios and D.~Sutherland,
  Phys.\ Rev.\ Lett.\  {\bf 114}, no. 7, 071601 (2015)
  doi:10.1103/PhysRevLett.114.071601
  [arXiv:1406.4422 [hep-th]].

  
\bibitem{kovtun} 
  P.~Kovtun and L.~G.~Yaffe,
  Phys.\ Rev.\ D {\bf 68}, 025007 (2003)
  doi:10.1103/PhysRevD.68.025007
  [hep-th/0303010].



\bibitem{kovtunlec} 
  P.~Kovtun,
  J.\ Phys.\ A {\bf 45}, 473001 (2012)
  doi:10.1088/1751-8113/45/47/473001
  [arXiv:1205.5040 [hep-th]].

  
\bibitem{rezzolla} L. Rezzolla and O. Zanotti, 
''Relativistic hydrodynamics'', Oxford University Press, (2013)

\bibitem{tong} David Tong, lectures on Kinetic theory\\
  Available at \url{https://www.damtp.cam.ac.uk/user/tong/kinetic.html}

\bibitem{brauner}
  T.~Brauner,
  arXiv:1910.12224 [hep-th].

  
\bibitem{higgs} 
  M.~E.~Peskin and D.~V.~Schroeder,
  ``An Introduction to quantum field theory,''



  \bibitem{kadanoff} L.Kadanoff and P.Martin, Annals of Physics {\bf} 24 419-469 (1963)

   
\bibitem{boulware}
  S.~Deser and D.~Boulware,
  J.\ Math.\ Phys.\  {\bf 8}, 1468 (1967).
  doi:10.1063/1.1705368


\bibitem{kaja} 
  A.~Czajka and S.~Jeon,
  Phys.\ Rev.\ C {\bf 95}, no. 6, 064906 (2017)
  doi:10.1103/PhysRevC.95.064906
  [arXiv:1701.07580 [nucl-th]].

  

\bibitem{surowka}
  D.~E.~Kharzeev, J.~Liao, S.~A.~Voloshin and G.~Wang,
  Prog.\ Part.\ Nucl.\ Phys.\  {\bf 88}, 1 (2016)
  doi:10.1016/j.ppnp.2016.01.001
  [arXiv:1511.04050 [hep-ph]]\\

  
 \bibitem{westerveldt} P.J. Westervelt, J. Acoust. Soc. Am. {\bf 29},934–939 (1957) 44 

\bibitem{gtstart} 
  G.~Torrieri,
  Phys.\ Rev.\ D {\bf 85}, 065006 (2012)
  doi:10.1103/PhysRevD.85.065006
  [arXiv:1112.4086 [hep-th]].



  \bibitem{lighthill} M.J.Lighthill, Proc. R. Soc. Lond. A 211 (1952)
\url{http://doi.org/10.1098/rspa.1952.0060}


  \bibitem{small}
C.~Loizides,
Nucl. Phys. A \textbf{956}, 200-207 (2016)
doi:10.1016/j.nuclphysa.2016.04.022
[arXiv:1602.09138 [nucl-ex]].
  

\bibitem{ghosts} 
  G.~Torrieri,
  arXiv:1810.12468 [hep-th].

  
\end{thebibliography}
\end{document}